\documentstyle[floats, aps, epsfig]{revtex}

\begin{document}

\twocolumn[\hsize\textwidth\columnwidth\hsize\csname @twocolumnfalse\endcsname

\title{ Gate-Controlled Electron Spin Resonance in a GaAs/AlGaAs Heterostructure }

\author{H.~W.~Jiang}
\address{Department of Physics and Astronomy, University of California at Los Angeles, Los Angeles, CA 90095}
\author{Eli Yablonovitch}
\address{Department of Electrical Engineering, University of California, Los Angeles, CA 90095-1594}

\date{\today}

\maketitle

\begin{abstract}
The electron spin resonance (ESR) of two-dimensional electrons is investigated in a gated GaAs/AlGaAs heterostructure.  We found that the ESR resonance frequency can be turned by means of a gate voltage.  The front and back gates of the heterostructure produce opposite g-factor shift, suggesting that electron g-factor is being electrostatically controlled by shifting the equilibrium position of the electron wave function from one epitaxial layer to another with different g-factors.
\end{abstract}

]

\narrowtext

Isolated electron spins in low temperature semiconductors are now recognized \cite{Loss} to have considerable potential for storing and manipulating quantum information. One of the great advantages of a spin in a semiconductor is that it can be embedded into a transistor structure, and it can thereby lend itself to large-scale integration of a quantum information processor. One essential element for spin-based quantum information processing is to be able to individually address the spins, or qubits.  In
an innovative paper, Kane \cite {Kane} proposed that the nuclear spin of a donor atom in Si can be manipulated and controlled, via the hyperfine interaction between the electron and nucleus, by a transistor gate.  We have recently suggested\cite{Vrijen} that this gate-controlled spin concept should be implemented directly on electron spins, since electronic band structure directly accesses the electron g-factor, whose matrix elements actually resemble those for effective mass.
In this case, the g-factor of an individual electron is tuned by a local gate electrode with respect to the frequency of a constant microwave field, to bring the spin in and out of the resonance.

There is a large body of work \cite {Hofmann} on the influence of composition and quantum well structure on g-factors.  Adjacent semiconductor heterostructure layers can have very different electron g-factors.  For example, Si-Ge alloys change from g=1.99 to g=0.82 over a narrow range of alloy composition.  Likewise GaAs has g=-0.44, while AlGaAs has g=+0.4.  Thus, the field induced shifting of the electron wave function between such layers can produce large g-factor changes, allowing direct g-factor
tuning by means of a gate voltage.

In this paper, we report our observations of gate-voltage tuned ESR in a two-dimensional electron system.  We demonstrate that the electrostatic field of a gate can effectively adjust the weighting of the electron wave function between heterostructure layers of different composition producing a large g-factor change.

The sample used for these experiments is a modulation doped GaAs/Al$_{0.3}$Ga$_{0.7}$As heterostructure.  The layers were grown by molecular-beam epitaxial on the $\langle$001$\rangle$ face of a GaAs wafer.  A 40nm undoped Al$_{0.3}$Ga$_{0.7}$As spacer layer was used to separate the Si donor layer (n=1x10$^{18}$/cm$^3$, 50nm thick) from the two-dimensional electron gas (2DEG) formed between the spacer and a 500nm GaAs buffer layer.  NiCr gates were evaporated both on the front and back of the sample.
Biasing the gate
allowed us to control both the electrical field perpendicular to the 2DEG plane and the density of the 2DEG.  To ensure good electrical insulation between the gate and the 2DEG another undoped Al$_{0.3}$Ga$_{0.7}$As layer (100nm) was included on top of the doped Al$_{0.3}$Ga$_{0.7}$As layer, followed by a 10nm GaAs cap layer.  Standard photo-lithography patterned a large area channel of width 150$\mu$m and length 450$\mu$m.  Indium was diffused into the channel to form Ohmic contacts.  The mobility of
the un-biased device at liquid helium temperature was 800,000cm$^{2}$/V-sec.

    To monitor electron spin resonance in bulk semiconductor systems, it is customary to detect microwave power absorption at spin resonance.  To obtain adequate signal amplitude, about 10$^{12}$ spins are normally required.  For our structure, there are only about 10$^{7}$-10$^{8}$ electrons available in the active channel. Therefore, we have chosen to detect the ESR by monitoring the electrical resistance of the source/drain channel.

It was demonstrated as early as 1983, in pioneering work by Stein, v. Klitzing, and Weimann \cite {Stein}, that the magnetoresistance of the 2DEG can be very sensitive to spin resonance, when the Fermi level is located between spin-split states of a given Landau level. Recent work \cite {Meisels,Vitalov,Dobers,Kane2,Wald,Kronmuller} on a variety of GaAs based devices have further demonstrated that resistive
detection is extremely effective for studying the magnetic resonance of electron as well as the nuclear spin.

Our experiment was carried out in a top-loading Helium 3 cryostat in a superconducting magnet.  A low-loss coaxial cable was used to deliver microwave radiation ($\approx$ 1mW) to the sample.  Fig. 1 illustrates the setup for detecting the ESR signal by means of source/drain channel resistance.  An ac probe current I$_{ac}$=200nA at 720Hz was applied from the source to the drain.  Then a lock-in amplifier monitored the channel resistance R$_{xx}$ through two additional electrical contacts along the
channel.
The microwave radiation, provided by a HP sweep generator was modulated at 100\% with a frequency of 10.8 Hz, much lower than the probe current frequency.  A second lock-in amplifier synchronous at 10.8Hz then measured the microwave induced change in resistance $\delta$R$_{xx}$.  This double modulation technique discriminated against possible thermo-voltaic or photo-voltaic effects.  The temperature for this experiment was chosen to be 1.2K, although lower temperatures were also studied.  At this
temperature over 70\% of the electron spins are already well polarized at a moderate field, about 2.5T.

\begin{figure} [tb]
\begin{center}
\epsfig{file = 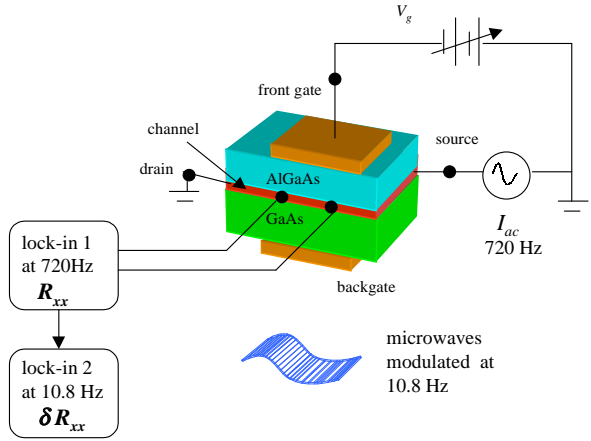, width = 8cm}
\end{center}
\caption {Diagram of the experimental setup for monitoring electron spin resonance and for controlling the spin orientation.}
\label{Figure 1}
\end{figure}

The experiment was carried out in the quantum Hall effect regime.  In Fig. 2, we show the typical channel resistance $\rho_{xx}$ versus magnetic field, and the corresponding change in channel resistance $\delta \rho_{xx}$ due to microwave radiation, all in Ohms as a function of perpendicular magnetic field on the 2DEG.  The carrier density is n$\approx$1.8x10$^{11}$/cm$^2$, with no dc voltage applied to the gate.  The oscillations in channel resistance can be roughly understood as the successive filling
of
Landau levels as the magnetic field is reduced.  The number of filled quantum states, the  filling factor $\nu$, is given by hn/cB where h is Planck's constant.  In this terminology, there are two spin states S$_{z}$=$\pm$1/2 for each Landau level.  For example, a filling factor $\nu$=3 indicates that both spin states of the N=0 lowest Landau level, and S$_{z}$=+1/2 of the next higher N=1 Landau level, are fully occupied by electrons as shown in the inset.  The majority of features displayed in the Fig.
2(b)
are not due to ESR.  Their origin has been commonly identified in the literature as being due to microwave heating. However, the sharp peak at B$\approx$2.5T is due to the ESR, whose position depends strongly on the microwave frequency.

We have worked mostly in a narrow range of gate voltage around V$_{g}$=+0.1V.  At this gate voltage, the density of the 2DEG is about 1.9x10$^{11}$/cm$^{2}$, corresponding to the S$_{z}$=+1/2 state of the second Landau level ({\it i.e.}, $\nu$=3, N=1) at a magnetic field of about 2.65T.  It is worth noting here that the ESR signal was indeed detected for several other odd filling factors ({\it i.e.}, $\nu$=1, 5, and 7).  We found that the ESR signal can be detected both in channel resistance, and gate
capacitance (or density of states).  The change in occupation density for the S$_{z}$ =-1/2 state at ESR confirms that there are actual spin flips in the sample.  The ESR linewidth ({\it i.e.}, full width at half-maximum) was found to be around 70-150G (corresponding 35-70MHz) depending on the excitation power and temperature.  The linewidth is known to be inhomogenously broadened by the hyperfine interactions with nuclear spins, Ga$^{69}$, Ga$^{71}$ and As$^{75}$ all having non-zero spin angular
momentum, I=3/2.

Fig. 2(b) shows that the electron spin state can control the channel resistance.  Now we will show that the gate can in turn control the spin.  In this part of the experiment, we varied the bias voltage around filling factor $\nu$=3 from 0.1V to 0.16V.  The ESR signal is best detected at exactly $\nu$=3, and the ESR signal strength diminishes quickly on either side away from full filling.  The gate voltage variation of 0.06V introduced a 12$\%$ density (or filling factor) change, about the limit of our
sensitive range.
Even within this rather small gate voltage range, we have been able to monitor the shift in ESR spectrum.

\begin{figure} [tb]
\begin{center}
\epsfig{file = 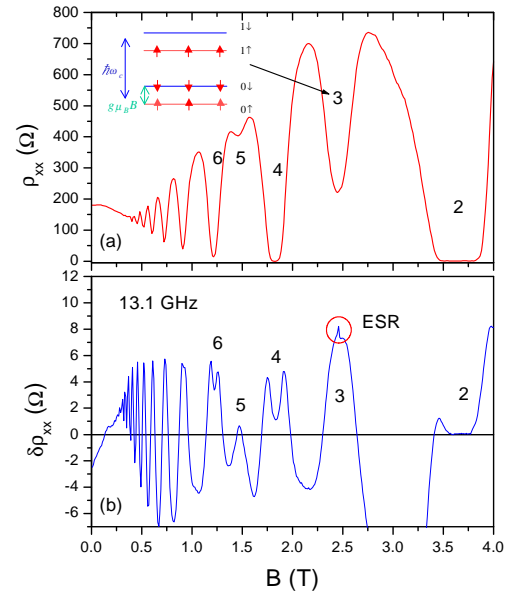, width = 7cm, clip=}
\end{center}
\caption{(a) Typical trace of the resistivity $\rho_{xx}$ as a function of the magnetic field.  Landau level filling factors $\nu$ are indicated.  Inset: energy diagram for the case of $\nu$=3. (b) The microwave radiation induced resistivity change $\delta$$\rho_{xx}$.  Note the ESR feature around $\nu$=3.}
\label{Figure 2}
\end{figure}

Fig. 3(a) shows a sequence of ESR spectra at different gate voltages.  At a fixed microwave frequency of 14.1GHz, the peak position shifts clearly and progressively from 2.672 to 2.682T as the gate voltage is increased.  The experimentally measured g-factor versus applied electric field E is plotted in Fig. 4.  Although the variation of the g-factor is only about 0.5$\%$, that tunes over 1 linewidth, within this voltage range.   In another sample from the same wafer, we have also placed a back-gate on
the GaAs substrate that is about 0.5mm away from the 2DEG.  The ESR spectra for different back-gate voltages is shown in Fig. 3(b).  However, in contrast to the front gate case, the peak position actually shifts towards lower fields with increasing positive gate voltage.  The g-factor has a ``blue shift" rather than a ``red shift" observed for the front gate case.

Both front and back gate voltages are measured with respect to the 2DEG channel, which is grounded.  Thus in both cases, a positive gate increases the Fermi Energy of the 2DEG.  The possibility of a 12$\%$ change of 2DEG density leading to a g-factor shift can be ruled out experimentally.  The increasing positive gate voltage would increase the density of the 2DEG for {\it both} the front and back gate cases.  In contradiction, a field shift in the opposite direction is observed for the back gate case!

\begin{figure} [tb]
\begin{center}
\epsfig{file = 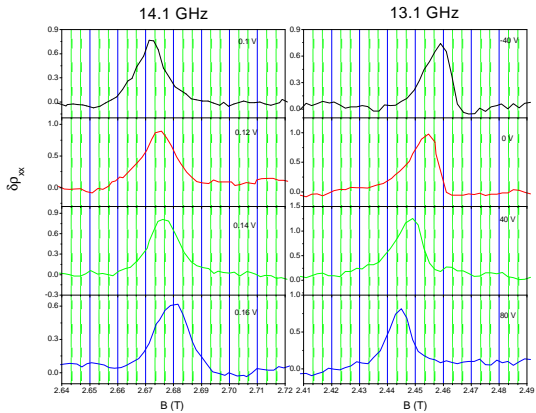, width = 8cm, clip=}
\end{center}
\caption{Electrically detected electron spin resonance spectra at a sequence of gate voltages for (a), a front gate and, (b), a back gate.  For the front gate case, the resonant peak moves progressively to higher magnetic field when the amplitude of the bias gate voltage is increased.  In contrast, the peak shifts towards lower field for the back gate case. (The smooth non-resonant background of the signal was subtracted for clarity.)}
\label{Figure 3}
\end{figure}

It is well known that the g-factor in a 2DEG system depends on magnetic field as well as Landau level index N as follows: g(B, N) = g$_{0}$-c(N+1/2)B, where g$_0$ and c are sample dependent constants.  In an earlier experiment, the g-factor was found to diminish continuously as magnetic field is increased \cite {Dobers2}.  This g-factor dependence was explained quantitatively by taking into account of the nonparabolicity of the bulk band structure \cite {Lommer}.  For the nearly parabolic bulk GaAs band,
the g-factor is known to be g=-0.44.  Note that this value deviates significantly from the free electron value of g=2.0023 due to spin-orbit coupling.  As the Fermi energy of the degenerate 2DEG increases, the energy band deviates progressively from the parabolic case, which leads to a reduction of the spin-splitting.  This nonparabolicity effect was indeed observed in our experiment (not shown) for large variations of B at a given Landau level. This nonparabolicity cannot however explain our
gate-controlled observations.
In the first place, the shift would be in the same direction for front and back gate cases, contrary to what is seen.  In the second place, the employed range of magnetic tuning field $\Delta$B should result in a g-factor change of only 3c$\Delta$B/2 or about 0.045$\%$ (for a typical value of c$\approx$0.014 T$^{-1}$) which is far less than the observed g-factor change of 0.5$\%$.

The mechanism that we invoke for the opposite g-factor shift between front and back gate cases is ``g-factor engineering" of the hetero-layers\cite {Vrijen}.  In this picture for the front gate case, as the magnitude of the gate voltage is increased, the wavefunction of the 2DEG is redistributed towards the Al$_{0.3}$Ga$_{0.7}$As side, as illustrated graphically in the inset of Fig. 4.  Since the g-factor of Al$_{0.3}$Ga$_{0.7}$As is about g=+0.4, the effective electronic g-factor, is consequently
reduced.  Since the energy barrier against wave function redistribution on the Al$_{0.3}$Ga$_{0.7}$As side is about 0.2eV, the change in effective g-factor is expected to be relatively modest, as observed.  For a back gate, an increasing gate voltage would enhance the weight of the wavefunction on the GaAs side, which increases the g-factor.  This wave function redistribution induced ESR is very different in origin, and is much greater than those due to due to g-factor anisotropy in different crystal
directions, that have previously observed \cite{Graeff}.

\begin{figure} [tb]
\begin{center}
\epsfig{file = 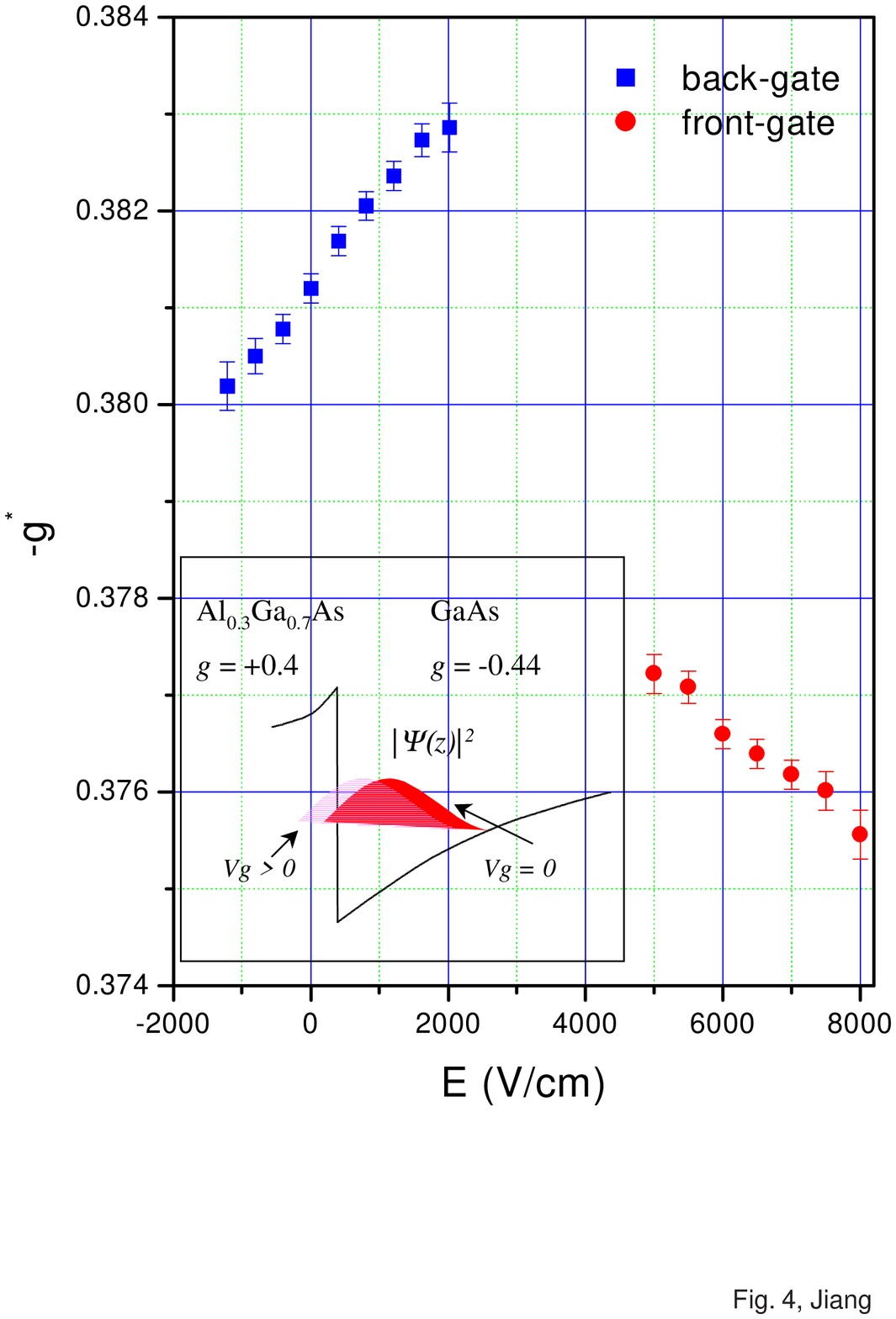, width = 8cm, clip=}
\end{center}
\caption{Experimentally determined electronic g-factor as a function of the applied electric field, for both the front and back gate.  The plotted electric field is simply the applied voltage divided by insulator thickness. (no attempt was made to include space charge self-consistently). Inset: The two-dimensional electrons are trapped in the ``triangle" shaped quantum well near the interface of the GaAs and Al$_{0.3}$Ga$_{0.7}$As materials. The electron wavefunction shifts back and forth for a positive
front gate bias voltage, and for V$_{g}$=0 .}
\label{Figure 4}
\end{figure}

To verify this wavefunction model quantitatively, we have performed a self-consistent calculation ({\it i.e.}, solving the Schodinger and the Poission equations of the band structure simultaneously) by using commercial semiconductor modeling software\cite {software}. The wavefunction distribution was evaluated for different front-gate bias voltages.  The g-factor was then determined by a weighted average over the two regions: $g=\bigl\lmoustache \big\arrowvert \psi (z)\big|^{2}g(z)dz$.  A g-factor shift
of about 0.7\% was obtained for a bias voltage difference of 60mV.  This simulation is in surprisingly good agreement with our experimental observations.  Although, this self-consistent calculation is intuitively informative, it is no substitute for a full band structure calculation.  Spin-orbit coupling, isotropic and anisotopic k-dependent contributions, etc., would be required to obtain good quantitative theoretical agreement.  For the application this effect to spin-based quantum information
processing, one requires the control of individual electrons, a far more challenging task.  However, we believe the demonstrated gate controlled ESR should be, in principle, applicable to the single spin case.

In conclusion, we have demonstrated in a GaAs/AlGaAs heterostructure the gate can control the electron spin by tuning it in and out of ESR resonance frequency.  Both red-shift and blue-shift of the ESR frequency were observed for positive front gate and positive back gate, respectively, proving that Fermi level changes cannot account for the g-shift.  The observations suggest that the gate controlled ESR is due to the tuning of the electron wave function probability weight between hetero-structure layers
of different compositional g-factor.

We would like to thank D. P. DiVincenzo and K. L. Wang for useful discussions. The work was supported by the Defense Advanced Research Projects Agency and Army Research Office under grant $\#$ MDA972-99-1-0017.

\end{document}